# Four-Channel WDM Graphene Optical Receiver

Laiwen Yu[1,+], Yurui Li[2,5,6,+], Hengtai Xiang[1], Yuanrong Li[1], Hengzhen Cao[1], Zhongyang Ji[2,6], Liu Liu[1,4], Xi Xiao[3], Jianbo Yin[2,5,6], Jingshu Guo[1,4,*], Daoxin Dai[1,4]

[1] *State Key Laboratory of Extreme Photonics and Instrumentation, College of Optical Science and Engineering, International Research Center for Advanced Photonics, Zhejiang University, Zijingang Campus, Hangzhou, 310058, China.*

[2] *State Key Laboratory of Advanced Optical Communications System and Networks, School of Electronics, Peking University, Beijing 100871, P. R. China.*

[3].*National Information Optoelectronics Innovation Center, China Information and Communication Technologies Group Corporation (CICT), Wuhan 430074, China.*

[4] *Jiaxing Key Laboratory of Photonic Sensing & Intelligent Imaging, Intelligent Optics & Photonics Research Center, Jiaxing Research Institute, Zhejiang University, Jiaxing 314000, China.*

[5] *Academy for Advanced Interdisciplinary Studies, Peking University, Beijing, 100871, China.*

[6] *Beijing Graphene Institute, Beijing 100095, P. R. China*

[+]*These authors contributed equally.*

*Corresponding Author: jsguo@zju.edu.cn*

**Abstract**

Silicon photonics with the advantages of low power consumption, low cost, and high yield is a crucial technology for facilitating high-capacity optical communications and interconnects. The graphene photodetectors (GPDs) featuring broadband operation, high speed, and low integration cost can be good additions to the conventional SiGe photodetectors, supporting silicon-integrated on-chip photodetection in new wavelength bands beyond 1.6 μm (e.g., U-band and 2 μm). Here we realize a silicon-integrated four-channel wavelength division multiplexing (WDM) optical receiver based on a micro-ring resonator (MRR) array and four p-n homojunction GPDs. These GPDs based on the photo-thermoelectric (PTE) effect operating under zero (current) bias exhibit responsivities of ~1.1 V W$^{-1}$ and flat frequency responses up to 67 GHz which is set-up limited. The GPDs show good consistence benefiting from the compact active region array (0.006 mm$^2$) covered by a single mechanically exfoliated hBN/graphene/hBN stack. Moreover, the WDM graphene optical receiver realized the 4×16 Gbps non-return to zero (NRZ) optical signal transmission. To the best of our knowledge, it is the first GPD-array-based optical receiver using high-quality mechanically exfoliated graphene and edge graphene-metal conduct with low resistance. Apparently, our design is also compatible with CVD-grown graphene, which can also result in a good consistence of the GPDs. This work shed light on the large-scale integration of GPDs with high consistency and uniformity, enabling the application of high-quality mechanically exfoliated graphene, and promoting the development of the graphene photonic integrated circuits.

**Introduction**

The ever-growing demand of global data traffic is driving the development of next-generation optical communication and interconnection technologies, standards and modules such as transmitters and receivers[1,2]. Silicon photonics has been attended significant concerns due to its high integration density and low cost benefiting from the complementary metal-oxide-semiconductor (CMOS) fabrication process[2]. With the help of wavelength division multiplexing (WDM) technology, the silicon photonic transmitters and receivers have been widely used in optical interconnect in data centers and optical communications[2,3]. Currently, the silicon photonic optical receivers mainly working at O-band or C+L-bands, and typically use the SiGe photodetectors (PDs)[4]. Extending the optical communication bands is a direct and effective solution for capacity improvement[5], e.g., the U-band (1.625 μm-1.675 μm) communication has received extensive attention currently[6]. However, Ge has a cut-off wavelength of ~1.6 μm for efficient optical absorption[4]. Meanwhile, the integration of III-V PDs on Si needs bonding fabrication technology[7] or special integration technology[8] in cost of larger cost and lower yield. Therefore, the high-speed and CMOS-compatible silicon integrated PDs operating in a wide band (e.g., from near-infrared to mid-infrared) are in urgent demand. Fortunately, graphene can provide a promising solution for its ultrafast carrier dynamics[9], high carrier mobility[10], and broadband photoresponse[11]. Moreover, as a two-dimensional material, graphene can be integrated onto photonics platforms without lattice mismatch, and can even be compatible with CMOS back-end line processing[12].

Recently, large-bandwidth graphene photodetectors (GPDs) have been demonstrated using photoconductor or phototransistor structures based on PV effect[13], bolometric (BOL) effect[14-16], and photoconductive (PC) effect[14], the G-Si heterostructure based on IPE effect[17], and the p-n homojunction structures based on the photothermoelectric (PTE) effect[18-24]. The G-Si heterostructure PDs typically exhibit limited quantum efficiency, and photoconductors/phototransistors tend to suffer from substantial dark current (~mA scale under bias). In contrast, zero-biased p-n homojunction GPDs based on the PTE effect (PTE GPDs) have emerged as the preferred approach for high-speed GPDs development due to their superior signal-to-noise ratio. However, the sensitivity and linearity of the PTE GPDs still have a significant gap compared to Si-Ge PDs. Better quality graphene[25] and graphene-metal contacts[26] are urgently needed to improve the responsivity and reduce the thermal noise of the PTE GPDs.

The integration methods of graphene on photonic platforms have two ways. The CVD-grown graphene is usually used by wet-transfer[27] or semi-dry transfer[28] methods, supporting large-scale (e.g., wafer-scale[29]) photonic integration. In terms of the PTE PDs using CVD-grown graphene, the state-of-the-art works have demonstrated 105 Gbps of single-device direct detection speed[24], the bandwidth of >67 GHz[21], and the responsivity of 10 V W$^{-1}$ [20]. Alternatively, mechanical exfoliation can be used to transfer graphene from high-quality bulk graphite using adhesive tapes without water or solvent involved[27]. This method results in graphene with superior material quality compared to CVD-grown graphene, specifically exhibiting higher mobility and fewer impurities. Moreover, the mechanical exfoliated graphene can usually be encapsulated by hBN/graphene/hBN stack structure[18] to realize better stability and mobility of graphene, and the edge graphene-metal conduct can be realized with much lower resistance. In a representative work, a record responsivity of 90 V W$^{-1}$ [22] were reported using the stack of hBN/graphene/hBN integrated on MRRs. While CVD-grown graphene has been applied to construct several active photonic integrated circuits (PICs)[16,30], the photonic devices based on mechanical exfoliated graphene have only be used in single photonic device to the best of our knowledge due to the limited transfer area.

In this work, we propose and demonstrate an integrated four-channel WDM optical receiver based on MRR wavelength demultiplexing filters and PTE GPDs with mechanically exfoliated monolayer graphene. A large hBN/single layer graphene (SLG)/hBN heterostructure is transferred by the mechanical exfoliation method onto a specially designed active region array with total footprint of only 100×60 μm$^2$. The four PTE PDs are based on thin-silicon slot-waveguide as our previous work introduced[23]. When operating at 1550.25/1548.7/1546.4/1544.68 nm, the present GPDs achieve responsivities of ~1.1 V W$^{-1}$ and flat frequency response up to 67 GHz (set-up limited).

The linear dynamic ranges are measured from 0.01 mW to 0.34 mW under zero current. To the best of our knowledge, this is the first work using mechanically exfoliate graphene to realize arrayed GPDs, based on which a WDM optical receiver supporting 4×16 Gbps NRZ transmission is demonstrated. In this work, the wavelength band of ~1.55 μm is chosen for convenience of device test. This work can be easily extended to longer wavelengths (e.g., U-band, 2-μm-band and beyond).

**Results and discussion**

**Design and simulation.** Figure 1 illustrates the working principle of the proposed graphene-based optical receiver, which is composed of a 4-channel wavelength demultiplexing filter and four PTE GPDs. All the receiver is based on the 100-nm-thick silicon waveguide platform introduced in our previous work[23]. Light can be coupled into the fundamental TE mode of the waveguide through the fiber-to-chip grating coupler, and then demultiplexed by the four MRRs to channels #1-#4. The semi-inverse-designed (SID) method[31] was used to design the broadband, low loss, fabrication-friendly directional couplers, which are used to construct the high-performance MRRs with good fabrication tolerances. In a free spectral range (FSR), the MRRs of channels #1-#4 correspond to the drop-port central wavelengths of 1550.25 nm, 1548.7 nm, 1546.4 nm and 1544.68 nm, respectively. The design details of the MRRs are introduced in **Supplementary Note 1**.

For each channel #$x$, the optical signal is transmitted through a MRR and a multimode interferometer (MMI) with a power splitting ratio of 90%/10%. The 10% port is connected to an output grating coupler for optical coupling alignment. The 90% power optical signal is then input to the corresponding GPD #$x$. In each GPD, the injected optical $TE_0$ mode is converted to the fundamental quasi-TE mode of the slot waveguide where the active region locates by a strip-to-slot-waveguide mode converter. As Fig. 1a shows, $L_c$ and $L_g$ are the channel length and width of the GPD, respectively. To ensure all the four active regions can be covered by the mechanically exfoliated hBN/SLG/hBN stack, we designed a compact active region array with a footprint of <100×60 μm². The spacing between the adjacent GPDs are ~4.6 μm. All the GPDs share the same structure which is like our previous work[23]. In the thin-Si slot waveguide of the active region, two metal (Al/Au) pads respectively connect the left and right silicon regions, acting as the gate electrodes. The mechanically exfoliated hBN/SLG/hBN stack is positioned on top of the 10-nm-thick $Al_2O_3$ dielectric insulator layer that encases the Si waveguide. In this way, the doping level of the graphene-sheet can be modulated separately by the two gate electrodes. The one-dimensional edge contacts between the source and drain (Cr/Au) electrodes and graphene are formed.

The active region waveguide is optimized for efficient optical absorption in graphene, resulting in parameters of ridge width = 0.6 μm, gap = 0.15 μm. Figure 1b depicts the fundamental quasi-TE mode field simulated by COMSOL with effective refractive indices of 1.902−0.0043i at a wavelength of 1.55 μm, corresponding to a mode absorption coefficient of about ~0.15 dB μm$^{-1}$ [23,32]. Figure 1c shows the simulated light propagation filed indicating the strip-to-slot mode conversion process. The strip-to-slot-waveguide mode converter is designed by 3D-FDTD simulation (Lumerical Inc.) with a coupling loss of ~0.7 dB at 1.55 μm.

The simulation of PTE processing has been discussed in our previous work[14,23]. The channel length is set to $L_c$ = 4.5 μm, taking account the PTE responsivity and active region footprint control, while the channel width is set to $L_g$ = 30 μm based on the two-dimensional material (2DM) stack size. The resistance with varied gate voltage of a reference FET (field effect transistor) structure was fabricated along with similar technology and measured to get the resistance-gate-voltage relation, from which graphene parameters are extracted: the graphene mobility $\mu$ = ~5.06×10$^4$ cm$^2$ V$^{-1}$ s$^{-1}$, the minimum conductivity $\sigma_{min}$ = ~2.6 mS, and $\zeta$ = 1 μm (see **Supplementary Note 2** for more details). The Seebeck coefficient is given by $S = -\frac{\pi^2 k_B^2 T}{3e}\frac{1}{\sigma}\frac{d\sigma}{d\mu_c}$, where $e$ is the electron charge, $k_B$ denotes the Boltzmann constant, and $\mu_c$ is the graphene chemical potential. Figure 2b shows the calculated Seebeck coefficient as a function of the gate voltage with room temperature $T$ = 300K. The electron temperature distribution $T_e(l, y)$ was obtained by solving the heat diffusion partial differential equation[14,23], where $l$ is the lateral position along the graphene sheet. Figure 2b depicts the lengthwise-average electron temperature $\overline{T}_e(l)$ [given by $\overline{T}_e(l) = \frac{\int_0^L T_e(l,y)dy}{L_g}$] with $V_{G1} - V_{CNP} = -1$ V and $V_{G2} - V_{CNP} = 1$ V (or vice versa), and the highest temperature is ~317 °C when the input optical power is $P_{in} = 0.1$ mW at 1.55 μm. In this case, the input optical power of the slot waveguide active region is 0.087 mW given that the mode converter loss is 0.7 dB. The PTE photovoltage $V_{PTE}$ was then calculated by $V_{PTE} = \int S(l) \frac{d\overline{T}_e(l)}{dl} dl$. In this way, the photoresponse under different gate voltages can be simulated. Figure 2c shows the gate-voltage-dependent photovoltage responsivity ($R_V = V_{PTE}/P_{in}$) map under zero bias when $P_{in} = 0.1$ mW at 1.55 μm. The result illustrates a six-fold pattern originating from the typical PTE photoresponse[33,34]. When $V_{G1} - V_{CNP} = -1$ V and $V_{G2} - V_{CNP} = 1$ V (or vice versa), the responsivity ($R_V = 1.83$ V W$^{-1}$) is maximized.

**Experimental results.** The silicon waveguides were fabricated on a SOI wafer with a 100-nm-thick top-silicon layer and 3-μm-thick buried-oxide layer using the processes of Electron-beam lithography (EBL) and inductively coupled plasma (ICP) etching. The thin silicon slot ridge waveguides can enhance the evanescent field vertically and thus enhance the light-graphene interaction. Then, the gate electrodes (Al/Au) are fabricated by electron-beam evaporation (EBE) and lift-off processes, followed by the deposition of a ~10-nm-thick $Al_2O_3$ insulator layer. The hBN/SLG/hBN stack was transferred onto the active region array by van der Waals assembly technique. The graphene-metal (Cr/Au) edge conducts were fabricated by EBL and EBE (more details are given in Method and **Supplementary Note** 3). Figure 3a presents the optical microscope picture of the as-fabricated 4-channel WDM optical receiver fabricated based on MRR array and compact integrated PTE GPDs. Figure 3b shows the optical microscope picture and the scanning electron microscope (SEM) picture of the fabricated GPD array. For GPDs #1-#4, the channel length $L_c$ is 4.5 μm, and the channel width $L_g$ (i.e., the optical absorption length in active region) is ~30, ~30, ~28, ~27 μm for the GPD #1, #2, #3, #4. Figure 3c shows the fabricated MRRs for wavelength demultiplexing as well as a SEM picture of an add-drop MRR.

Figure 4a shows the measured transmission spectra for the fabricated MRRs. The ELs at the center wavelengths for the four channels are <1 dB with ~1.5-nm channel spacing. In an FSR, all four channels have a 1-dB bandwidth of ~0.5 nm and 3-dB bandwidth of ~0.87 nm. The better performance of the MRRs contributes to the stability of the work of this system. The static performance of the as-fabricated GPDs was measured with three source meters[23]. Figures 4b and 4c present the measurement results of GPD #1 with $L_c$ = 4.5 μm and $L_g$ = ~30 μm. The resistance of the graphene channel under different gate voltages was measured under near-zero bias voltage of 1 mV, showing that the charge neutrality point voltage $V_{CNP}$ is about −1 V, as shown in Fig. 4b. The optical power input a GPD is denoted by $P_{in}$ which is extracted by the output power of fiber and the passive device losses (the MRR and the MMI). Figure 4c gives the measured $V_{PTE}$ under zero current at the wavelength of 1550.25 nm with $P_{in}$ = ~0.1 mW as a function of the gate voltages ($V_{G1}$, $V_{G2}$). Figure 4d shows the measured relation of $V_{PTE}$~$P_{in}$ under zero current for the four GPDs. By fitting the relation of $V_{PTE}$~$P_{in}$, the GPD #1 has a responsivity $R_V$ of ~1.2 V W$^{-1}$ when $V_{G1}$ = 2.1 V, and $V_{G2}$ = 7.2 V, and a linear dynamic range is 0.001-0.34 mW. The four GPDs have similar responsivities: GPD #2: $R_V$ = ~1.1 V W$^{-1}$ when $V_{G1}$ = 1.3 V, and $V_{G2}$ = 7.1 V; GPD #3: $R_V$ = ~1.3 V W$^{-1}$ when $V_{G1}$ = 5.2 V, and $V_{G2}$ = 7 V; GPD #4: $R_V$ = ~1 V W$^{-1}$ when $V_{G1}$ = 6.2 V, and $V_{G2}$ = 1.5 V. More details of the static performances of GPDs #2-#4 are given in **Supplementary Note 4**.

As shown in Fig. 5a, the frequency responses of the GPDs were measured by using a commercial light-wave component analyzer (LCA). The high frequency response of the GPD under zero-current bias was obtained from the measured $S_{21}$ calibrated by the responses of a RF amplifier and a RF probe. The modulated optical signal was generated by the LCA and was amplified with an erbium-doped fiber amplifier (EDFA). For each WDM channel, the signal was then injected to the GPD through the passive optical devices. Here a bias-Tee was used to apply the zero-current bias and collected the RF response simultaneously. Since the corresponding passive MRR is involved in the reference calibration path with help of the broadband MMI power splitter and the grating couplers, the influence of the narrow-band MRR can also be calibrated. Under an open circuit test mode (zero-current bias), the measured high frequency response of the all four GPDs are shown in Fig. 5b with good consistence, indicating that the 3-dB bandwidths of all the four GPDs are >67 GHz and beyond the setup measurement range. The self-driven GPDs with large bandwidth features low power consumption and high-speed data reception. More details of the high frequency response measurements are given in **Supplementary Note 5**.

Figure 6a depicts the measurement system of the high-speed optical data transmission, which includes an arbitrary waveform generator (AWG, M8195A), a commercial 40 GHz optical modulator, our WDM optical receiver, and a real-time oscilloscope (DSOZ594A). For each channel, the non-return-to-zero (NRZ) on-off-keying (OOK) signal generated by the AWG was amplified and injected to the optical modulator. The continuous wave light from tunable laser was modulated, amplified, and input to our receiver chip. The output signal of the GPD was amplified and then measured by the real-time oscilloscope. The time-domain signal is directly used to plot the eye diagram with no system optimization algorithm used. All the eye diagrams of the four channels are shown in Fig. 6b with date rate of 16 Gbps. Under zero-current bias, each GPD operated with optimized gate voltages applied to maximize the responsivity. The modulated optical power into each GPD is $P_{in}$ = ~2 mW. The carrier wavelengths for four channels are respectively 1550.25 nm, 1548.7 nm, 1546.4 nm, and 1544.68 nm.

The static and high frequency performance of the four GPDs show good consistency, which can be contributed to the single 2DM stack strategy using a compact active region array. The 3dB-bandwidths >67 GHz is comparable to the state-of-the-art works[21,24]. The MRRs of 0.5 nm bandwidth can support >60 Gbps data transmission theoretically. Therefore, we argue that the data transmission speed is limited by the responsivities of the GPDs. The ridge waveguide may cause strain in graphene, which degrades the PTE photoresponse. In future, more broadband WDM filters can be used, and the waveguide with flat top-interface can be used to avoid the strain.

**Conclusion**

In conclusion, we demonstrated a 4-channel WDM graphene optical receiver based on the MRR array and PTE GPDs operating at zero current. The MRRs realized a homogeneous spectral transmission at the bandwidth of 100 nm with ~1.5-nm channel spacing. The active region array of the GPDs is specially designed within a compact footprint of ~0.006 mm$^2$, enabling the cover of a single high-quality mechanically exfoliated hBN/SLG/hBN stack. In this way, the GPDs show good consistency of the experimental performance. Specifically, the four GPDs exhibit similar responsivities of ~1.1 V W$^{-1}$ with flat frequency responses up to 67 GHz which is set-up limited. Using our graphene optical receiver, the transmission of 4×16 Gbps NRZ optical signal is demonstrated with carrier wavelengths of 1550.25 nm, 1548.7 nm, 1546.4 nm, and 1544.68 nm. This work can be easily extended to longer wavelengths, e.g. U-band, 2-μm-band or even beyond. To the best of our knowledge, this work represents the inaugural application of mechanically exfoliated high-quality graphene in the scale integration of GPDs, being a step forward in the application of graphene photonic devices in optical communications and interconnects. Apparently, our design is also compatible with CVD-grown graphene, whose compact active region array can result in an improved consistency of GPDs in contrast to the conventional designs. In the future, the waveguide platform with a flat top-interface can be employed to eliminate the impact of graphene strain on the responsivities of PTE GPDs, thereby enabling higher-speed data reception.

**Methods**

**Transceiver Fabrication.** The thin silicon platform with 100-nm-thick Si layer was obtained from a standard SOI wafer with a 220-nm-thick top Si layer by using thermal oxidation and buffered oxide etch (BOE). The passive ridge silicon waveguides and circuits were fabricated by the processes of EBL and ICP dry-etching. For each GPD, the silicon slot ridge waveguide has a ridge width of 1.5 μm, a slot width of ~120 nm, an etching depth of 100 nm. Next, the 90-nm-thick aluminum (30 nm) /Aurum (60 nm) gate-electrodes were fabricated by EBL, EBE and lift-off in acetone, forming ohmic contacts with two parts of the silicon slot waveguide separately. Then, a 10-nm-thick $Al_2O_3$ layer was deposited on the slot ridge waveguide by the atomic-layer deposition (ALD) process. The hBN/SLG/hBN stack was transferred onto the chip. Then, the Cr (5 nm)/Au (150 nm) electrodes were fabricated by EBL, EBE and lift-off processes, forming the edge contact between graphene and the source/drain electrodes. Finally, the hBN/SLG/hBN stack is patterned by the EBL and RIE etching processes to isolate the adjacent GPDs electrically.

**Transfer Process of Graphene.** The hBN/SLG/hBN stack was fabricated by van der Waals assembly technique. The graphene and hBN flakes were exfoliated from highly oriented pyrolytic graphite and hBN crystals using Stoch tape and heat release tape respectively. The graphite and hBN crystals were purchased from HQ Graphene Company. The hBN/SLG/hBN stack was picked up by a stamp composed of a polycarbonate (PC) layer supported by a polydimethylsiloxane (PDMS) block. For more details on the van der Waals assembly technique, please refer to ref.[35]. The assembled stack was released onto the active region array after oxygen plasma treatment.

## Data availability

Data supporting the findings of this study are available from the corresponding authors upon reasonable request.

**Acknowledgements**

This work was supported by National Key R&D Program of China (2022YFA1204900); National Science Fund for Distinguished Young Scholars (61725503); National Natural Science Foundation of China (NSFC) (61905210, 61961146003, 91950205, T2188101); Zhejiang Provincial Natural Science Foundation (LR22F050001, LD22F040004); Leading Innovative and Entrepreneur Team Introduction Program of Zhejiang (2021R01001); The Fundamental Research Funds for the Central Universities. We thank Westlake Center for Micro/Nano Fabrication and ZJU Micro-Nano Fabrication Center for the facility support and technical assistance. We are grateful to the Polytechnic Institute of Zhejiang University for providing support with test equipment.

**Author contributions**

J. G. and L. Yu conceived the optical receiver. L. Yu designed the MRRs and the GPDs under guidance of J. G.. L. Yu fabricated the optical receiver. Yurui Li and J. Y. performed the transfer and characterization of the hBN/SLG/hBN stack. L. Yu and J. G. did the device characterization and system testing with help of Yuanrong Li. X. X.

contributed to the high frequency measurement. J. G. and L. Yu analyzed and discussed the experimental data. L. Yu prepared the first draft. J. G., L. Yu and D. D. completed the paper with input from all authors. J. G. and D. D. supervised and coordinated all the work.

**Competing interests**

The authors declare no conflict of interest.

**Additional information**

**Supplementary information** Design of the micro-ring resonators; Details on graphene characterization; Details on the fabrication of the graphene optical receiver; Static performances of GPDs #2, #3, #4; Experimental setup for the high frequency response measurements.

Correspondence and requests for materials should be addressed to J. G..

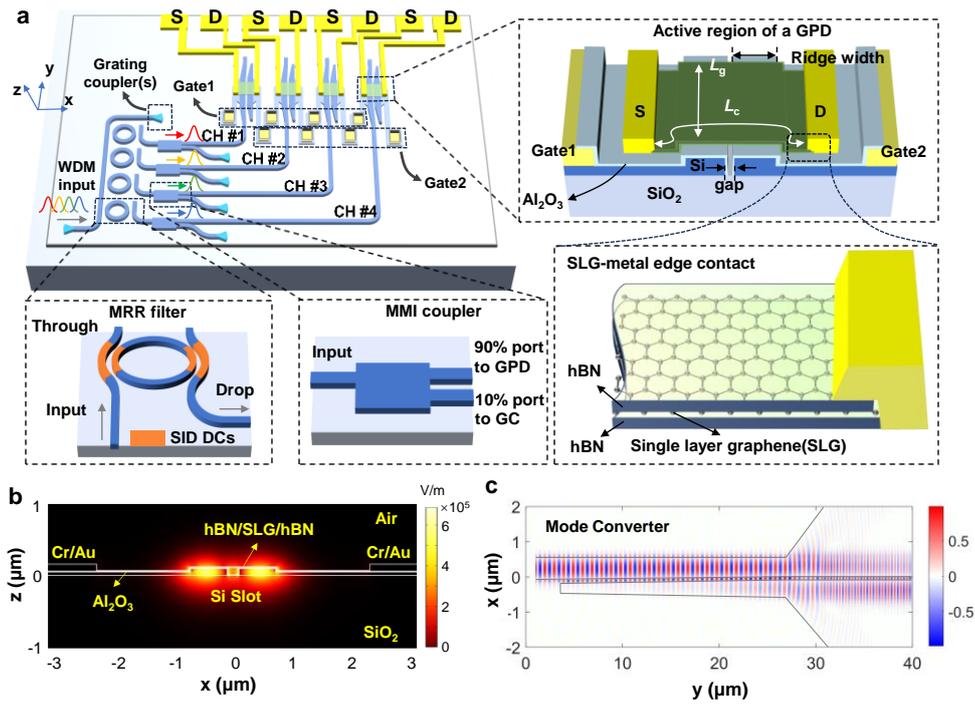

**Fig. 1 Design of the present four-channel WDM optical receiver based on GPDs. a** Schematic configuration of the photonic integrated circuit with grating couplers, MRRs, MMIs and the GPDs. Inset: Structures of a MRR, a MMI, the active region of a GPD, and the SLG-metal edge contact. $L_c$, channel length; $L_g$, Channel width; SLG, Single layer graphene; CH, channel; SID, semi-inverse design; DC, directional coupler. **b** Electric field distribution of the fundamental quasi-TE mode for the optimized slot ridge waveguide. **c** Normalized simulated light propagation in the mode converter when operating at 1.55 μm.

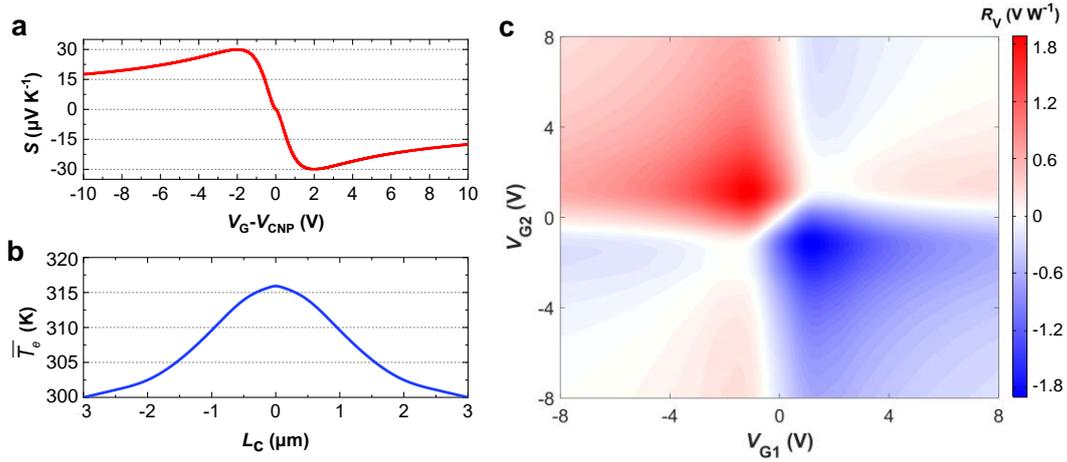

**Fig. 2 Simulation results of the PTE process. a** Calculated Seebeck coefficient as a function of gate voltage. **b** Calculated lengthwise-average electron temperature profile along the graphene channel between the source and drain contacts ($P_{in}$=0.1 mW). **c** Six-fold photovoltage responsivity map of the designed GPD at 1.55 μm with $L_c$=4.5 μm and $L_g$=30 μm, $P_{in}$=0.1 mW. $P_{in}$: input optical power to the GPD.

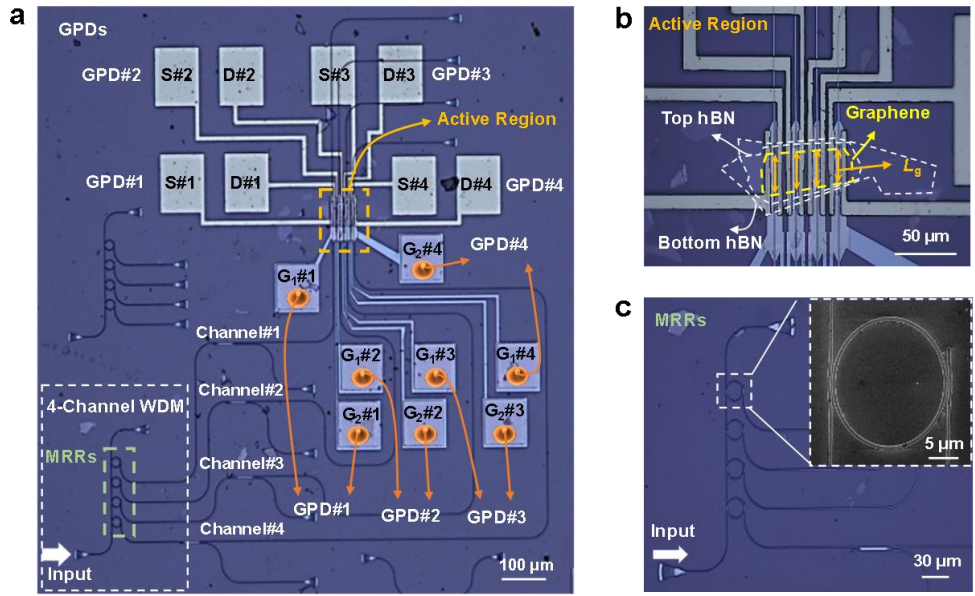

**Fig. 3 The fabricated 4-channel WDM optical receiver based on PTE GPDs. a** Microscope picture of the photonic integrated circuit. S#$n$, D#$n$ are respectively the source, drain electrodes of GPD #$n$ ($n$ = 1, 2, 3, 4). Gates are contacted through metal contacts ($G_1$#1-#4 and $G_2$#1-#4). Scale bar, 100 μm. **b** Microscope picture of the GPDs with $L_c$ = 4.5 μm and $L_g$ = ~30, ~30, ~28, ~27 μm (GPDs #1-#4). Scale bar, 50 μm. **c** Microscope picture of the MRR array. Scale bar, 30 μm. Insert figure, SEM picture of an add-drop MRR. Scale bar, 5 μm.

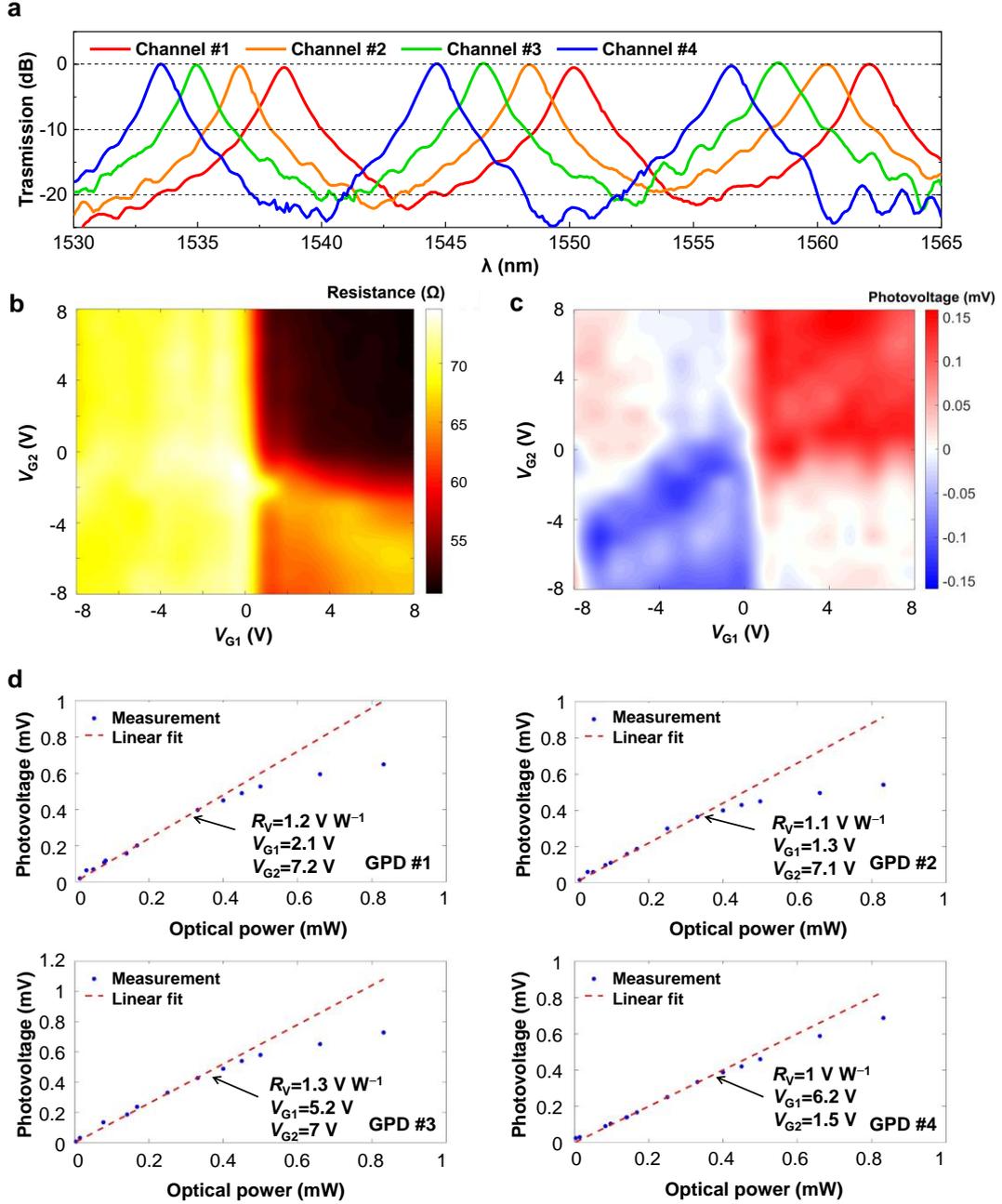

**Fig. 4 Static experimental results of the MRRs and the GPDs. a** Measured transmission spectrum of the fabricated MRRs. **b, c** Measured results of GPD #1 ($L_c$ = 4.5 μm and $L_g$ = ~30 μm) including the resistance map **b** of as a function of the gate voltages $V_{G1}$ and $V_{G2}$ and the photovoltage map **c** at 1550.25 nm with $P_{in}$ = ~0.1 mW under zero-current bias. **d** Photovoltage $V_{PTE}$ as a function of input optical power $P_{in}$ under zero-current bias with the gate voltages $V_{G1}$ and $V_{G2}$. Dots and dashed lines are respectively the measured results and fitting results governed by $R_V = V_{PTE}/P_{in}$.

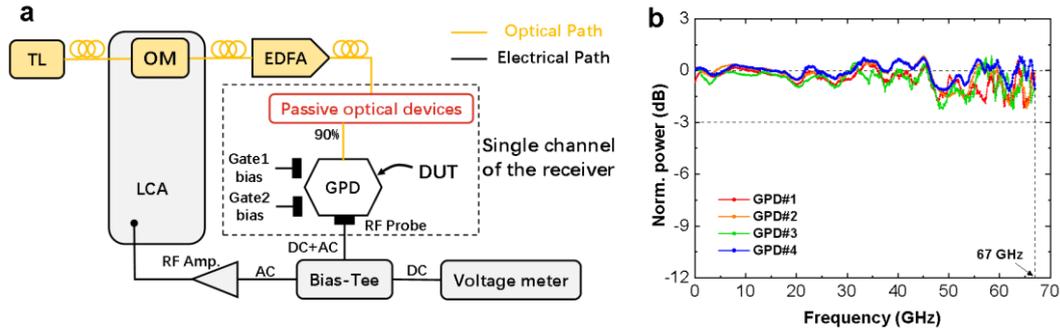

**Fig. 5 High frequency response of the GPDs. a** Measurement setup. **b** Normalized high frequency response of the four GPDs with zero-current bias. LCA: light-wave component analyzer, OM: optical modulator, RF Amp.: radio frequency amplifier, EDFA: erbium-doped fiber amplifiers, DC: direct current, AC: alternating current. Passive optical devices include the grating coupler, MRR, and MMI.

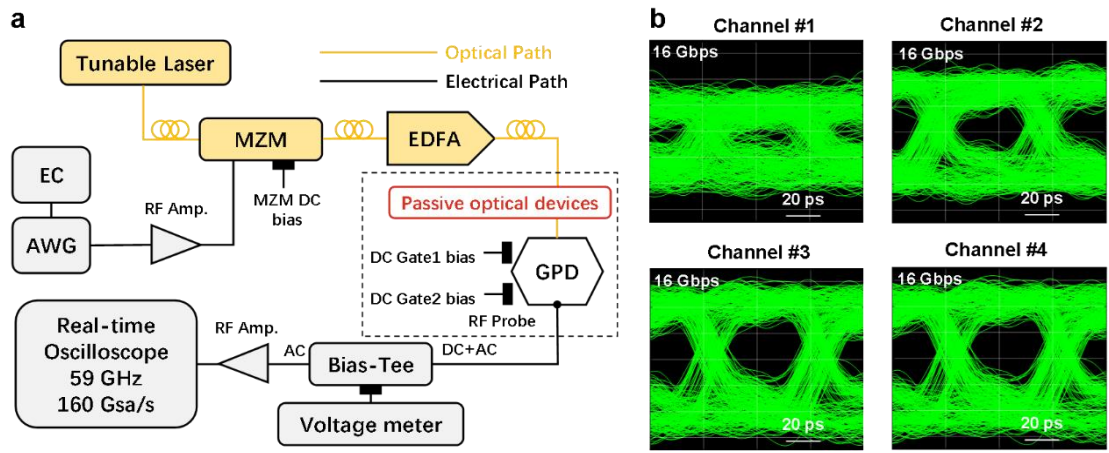

**Fig. 6 Demonstration of WDM optical data transmission by our optical receiver. a** Measurement setup. EC: Embedded Controller, AWG: arbitrary waveform generator, MZM: Mach–Zehnder modulator. **b** Eye diagrams (0.01 mV div$^{-1}$) of the 16 Gbps NRZ OOK data transmission by using four channels #1-#4, operating at zero-current bias. The modulated optical power input to the GPD is $P_{in}$=2 mW.

# Supplementary Information

**Supplementary Note 1. Design of the micro-ring resonators.**

The micro-ring resonator (MRR) array forms a 4-channel wavelength demultiplexing filter with a waveguide height of 100 nm. Figure S1a depicts the structure of one MRR, including a pair of Bezier-shaped couplers[1]. The transmission spectra of each MRR are obtained by using the transfer matrix method based on the phase shift of the ring and the transmission spectra of the couplers. The couplers are designed by the semi-inverse design (SID) method introduced in our previous works[1,2] using the finite-difference time-domain (FDTD) method. Since the two couplers (denoted by the orange waveguides in Fig. S1a) of the add-drop MRR are symmetric, one coupler is used to illustrate the design process.

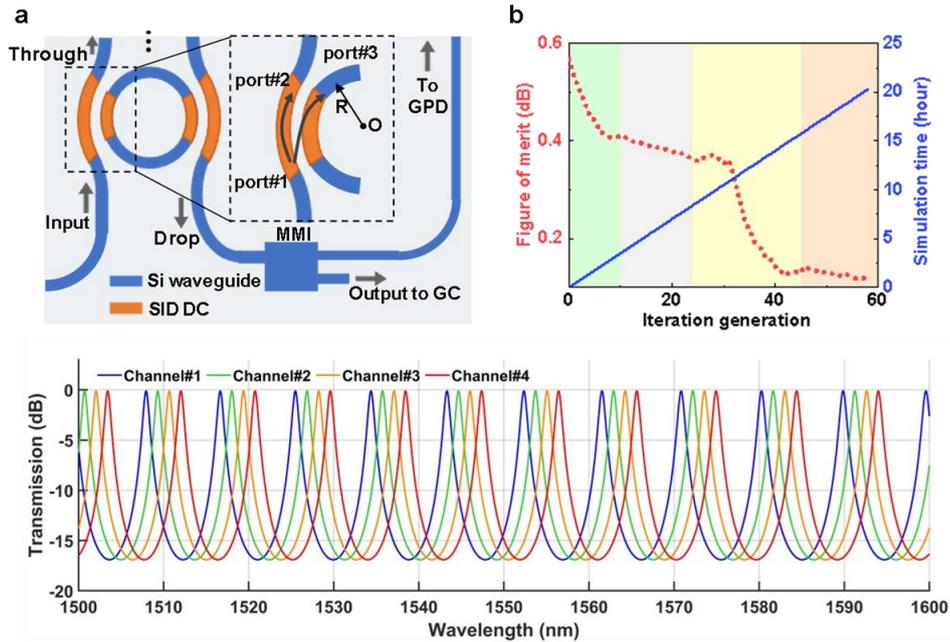

**Figure S1 Design of the micro-ring resonators. a** The MRR structure and the definition of the couplers. **b** Figures of merit as a function of the iteration generation for the couplers with design wavelength range of 1.5-1.6 μm. **c** The simulated drop-port transmission spectra of the designed four-channel MRRs.

The optimization figure of merit (*FOM*) for specifically wavelength $\lambda_i$ is defined by[1]:

$$FOM_{\text{in}}(\lambda_i) = -10\log_{10}\{1 - \sqrt{[CR - |S_{31}(\lambda_i)|^2]^2 + [1 - CR - |S_{21}(\lambda_i)|^2]^2}\}, \quad (S1)$$

where $S_{xy}$ is the coupling coefficient from mode #$y$ to mode #$x$, and *CR* is the target power coupling ratio. Here modes #1-#3 respectively refer to the $TE_0$ modes of ports #1-#3. To design a broadband directional coupler for the MRR, multiple operating wavelengths were considered:

$$FOM = \frac{1}{n}\sum_{i=1}^{n} FOM_{\text{in}}(\lambda_i). \quad (S2)$$

We choose five wavelengths ($n = 5$) as {1.5, 1.525, 1.55, 1.575, 1.6} μm and *CR* = 0.25 to make the MRR work well in a broadband. As Figure S1b shows, the *FOM* varies from 0.57 to 0.11 within 59 generations in the three stages of optimization, taking ~20 hours totally. By fine tuning the ring radius, the four MRRs were designed with the same free spectral range (FSR) and different resonance wavelengths. Figure S1c shows the simulated drop-port transmission spectra of the four-channel MRRs, each of which shows a 1-dB bandwidth of ~0.5 nm. As Fig. S1a shows, the optical signal output from drop port was guided to the multimode interferometer power splitter with 90%/10% splitting ratio.

**Supplementary Note 2. Details on graphene characterization.**

As shown in Figure S2, we performed a Raman Spectrometer (LabRAM HR Evolution, HORIBA, Japan, wavelength 532 nm, power<1 mW) and an atomic force microscopy (AFM, dimension ICON, Bruker, Germany) to monitor the SLG quality at the position of the final device. A typical Raman spectrum before further processing of the stack is shown in Figure S2a. The position of the combined hBN $E_{2g}$ peaks[3] from top and bottom flakes is Pos($E_{2g}$) ~1365 cm$^{-1}$ with a full-width half maximum FWHM($E_{2g}$) of ~9 cm$^{-1}$, as expected considering the top flake is bulk and that the planar domain size is limited by the flake size[4]. Pos(2D) ~2683 cm$^{-1}$, FWHM(2D) ~20.1 cm$^{-1}$, Pos(G) ~ 1582 cm$^{-1}$, and FWHM(G) ~ 12 cm$^{-1}$ confirm the presence of SLG. The AFM images of the slot waveguide and the active region with 2DM stack are shown in Fig. S2b and Fig. S2c, respectively. It can be seen that the 2DM stack atop the ridge slot waveguide is not a perfect conformal fit. Nevertheless, the maximum undulating height in the active region (~132 nm, in Fig. S2c) and the passive waveguide area (~135 nm, in Fig. S2b) is basically the same, indicating that the stack fits on the waveguide well.

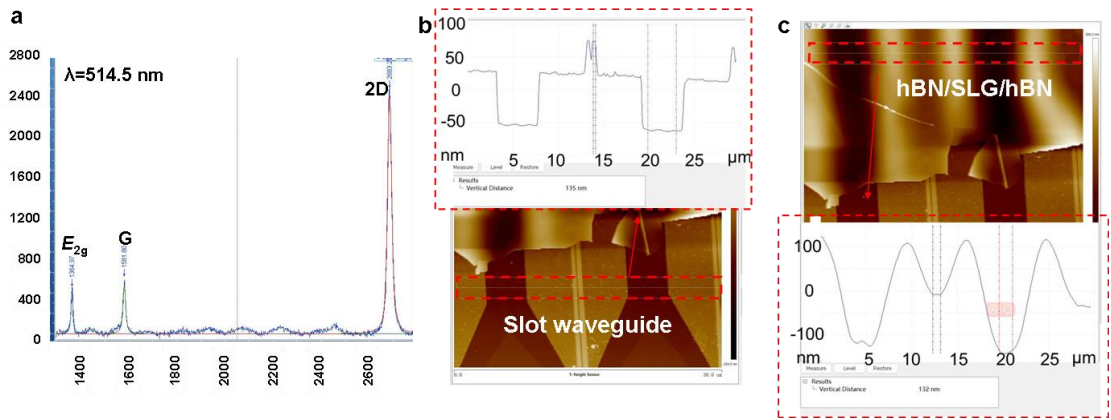

**Figure S2 Characterization of the 2DM stack. a** Raman spectrum measured at the position of the final device. **b** AFM image of the slot waveguide. **c** AFM image of the hBN/SLG/hBN atop the waveguide.

A reference field effect transistor (FET) was fabricated along with the graphene PDs, consisting of hBN/SLG/hBN channel with $L_c$ = 4.5 μm and $L_g$ = 30 μm. This FET shares the same 10-nm-thick $Al_2O_3$ and 15-nm-thick bottom hBN insulator layer with the graphene PDs. The total resistance between the drain and source electrodes ($R_{tot}$) was measured as gate voltage ($V_G$) varies, as shown in Fig. S1. As introduced in our previous work, the total resistance of the graphene can be expressed as:

$$R_{tot} = R_c + R_g = R_c + \frac{L_c}{L_g}\sigma^{-1}, \tag{S3}$$

where $R_c$ is the total resistance of the metal-graphene contacts, $R_g$ is the resistance of gate-tunable graphene, $\sigma$ is the electrical conductivity of graphene. The graphene electrical conductivity $\sigma$ can again be expressed by

$$\sigma = \sqrt{\sigma_{min}^2 + [\mu C_G(V_G - V_{Dirac})]}, \tag{S4}$$

where $\sigma_{min}$ is graphene minimum conductivity, $\mu$ is the carrier mobility of graphene, $V_{Dirac}$ is the gate voltage corresponding to the charge neutral point of graphene, and $C_G$ is the gate dielectric capacitance given by[4,14] $C_G = (C_{Al_2O_3}^{-1} + C_{hBN}^{-1})^{-1} = 8.83 \times 10^{-4}\ F\ m^{-2}$. By fitting the measured $R_{tot}$-$V_G$ curve with Eqs. S3 and S4, we have $R_c$ = ~65 Ω, $\sigma_{min}$ = ~2.6 mS, $\mu$ = ~5.06×10$^4$ cm$^2$·V$^{-1}$·s$^{-1}$ and $V_{Dirac}$ = −1 V. The fitting curve is shown by the red solid line in Figure S3.

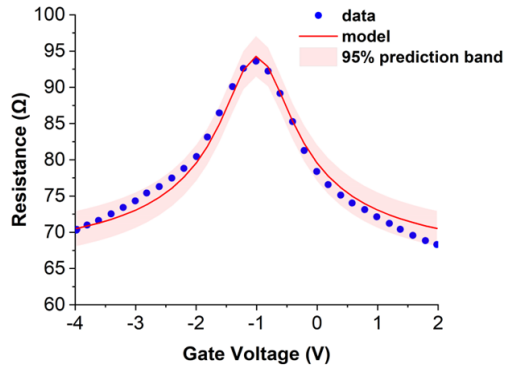

**Figure S3 Electrical characterization of a reference field effect transistor (FET).** Blue dots: measured data, solid line: fitting result.

**Supplementary Note 3. Details on the fabrication of the graphene optical receiver.**
Figure S4 illustrates the fabrication process of our graphene optical receiver. Firstly, the SOI with 100 nm silicon top layer was obtained from a standard SOI wafer with a 220-nm-thick top silicon layer by using thermal oxidation and buffered oxide etch (BOE) processes. The full-etched and shallow-etched Si photonic waveguides are defined by two round of electron beam lithography (EBL) and inductively-coupled plasma dry-etching (ICP) processes, as shown in Fig. S4a. The fabrication of the GPD active regions is shown in Figure S4b-f. The 90-nm-thick Al (30 nm)/Au (60 nm) gate-electrodes were fabricated by processes of EBL, electron-beam evaporation (EBE), and lift-off in acetone, forming ohmic contacts with two parts of the silicon slot waveguide separately. Then, a 10-nm-thick Al$_2$O$_3$ layer was deposited on the slot ridge waveguide by the atomic-layer deposition (ALD) process (Fig. S4c). The hBN/SLG/hBN stack was transferred onto the active region area (Fig. S4d). The 2DM stack was patterned by the EBL and RIE processes, followed by the fabrication of Cr (5 nm)/Au (150 nm) source and drain electrodes, in this way, the metal-SLG edge contacts were formed (Fig. S4e). Finally, the part of the 2DM stack between the adjacent GPDs was removed by oxygen plasma etching to ensure that each GPD works independently.

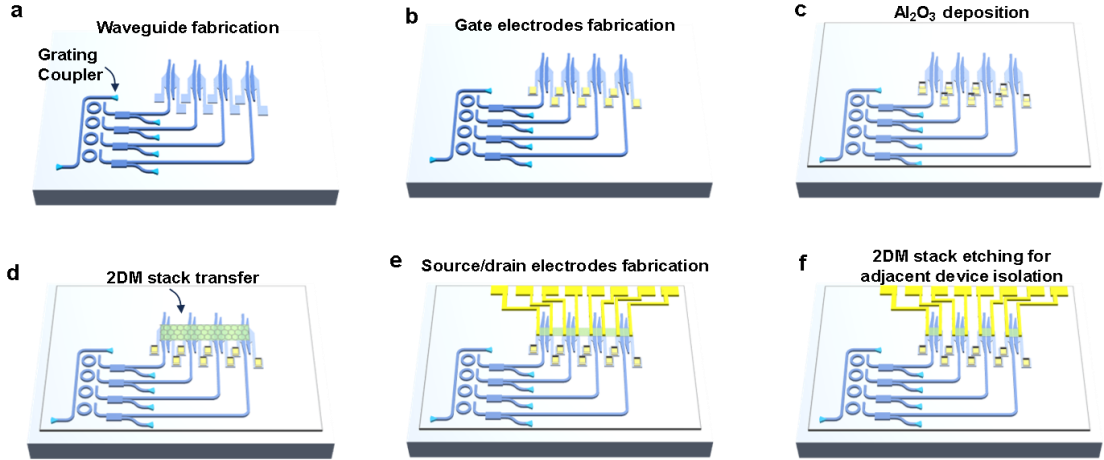

**Figure S4 Fabrication process of the graphene optical receiver. a** Fabrication of the Si full-etched and shallow-etched waveguides. **b** Al/Au gate electrodes fabrication by EBL, EBE, and lift-off processes. **c** Deposition of the Al$_2$O$_3$ dielectric layer. **d** hBN/SLG/hBN stack transfer. **e** Cr/Au-SLG edge conduct forming. **f** The oxygen plasma etching of the hBN/SLG/hBN stack for electrical isolation of the adjacent GPDs.

**Supplementary Note 4. Static performances of GPDs #2, #3, #4.**

As shown in Fig. S4a-c, the resistance of the graphene channel under different gate voltages was measured under near-zero bias voltage of 1 mV, showing that the charge neutrality point voltage $V_{CNP}$ is about −1 V. Figure S4d-f give the measured $V_{PTE}$ under zero bias at the wavelength of 1548.7 nm, 1546.4 nm, 1544.68 nm with $P_{in}$ = ~0.1 mW as a function of the gate voltages ($V_{G1}$, $V_{G2}$) for GPDs #2, #3, #4.

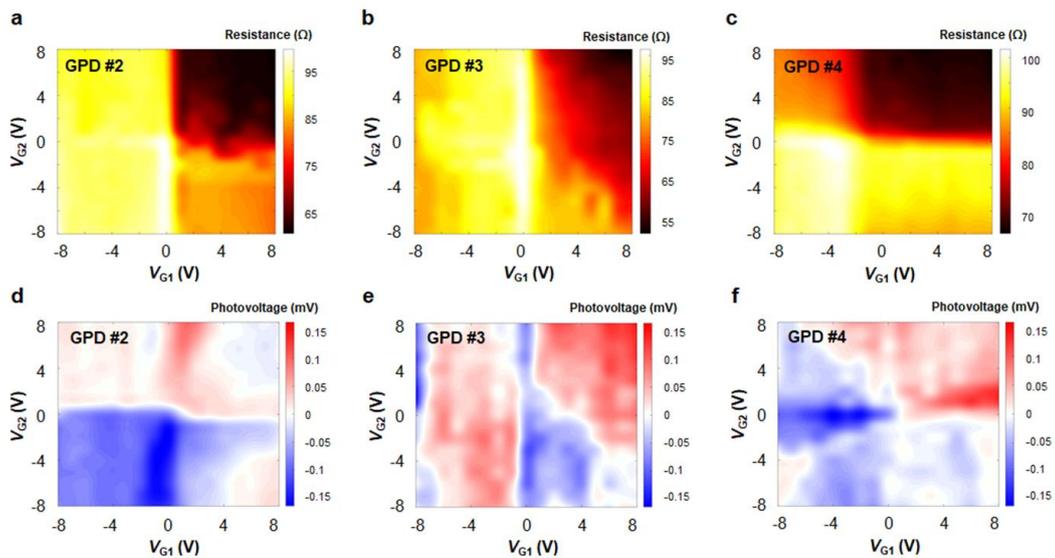

**Figure S5 Experimental static performances of GPDs #2, #3, #4. a-c** Measured resistance map of the fabricated

PTE GPD #2 **a**, #3 **b**, #4 **c** as a function of the gate voltages $V_{G1}$ and $V_{G2}$; **d-f** Measured photovoltage maps at 1548.7 nm, 1546.4 nm, 1544.68 nm with $P_{in}$ = ~0.1 mW under zero-current bias for GPD #2 **d**, #3 **e**, #4 **f**.

**Supplementary Note 5. Experimental setup for the high frequency response measurements.**

Figure S6 shows the experimental setup for frequency response measurements. Firstly, we calibrated the bias-Tee and cables in the test system. The optical signals generated by a tunable laser (TL) at $\lambda_{CH\#n}$ ($\lambda_{CH\#1,2,3,4}$ = 1550.25 nm, 1548.7 nm, 1546.4 nm and 1544.68 nm) was modulated by an optical modulator integrated in a light-wave component analyzer (LCA, KEYSIGHT N4373E). The modulated optical signal was then amplified by an EDFA and coupled to the grating coupler. In each channel, the light passed through a power splitter (90%/10%) with a multimode interferometer (MMI) structure. The 10%-power optical signal was given to the commercial photodetector (integrated in the LCA), forming a reference calibration path. In this way, the frequency responses of the MRR and optical modulator can be calibrated. The AC photocurrent signal generated by the GPD was then extracted by a bias-Tee (SHF BT 65-B) and amplified by an RF amplifier (SHF 807C). Then, the LCA received the amplified AC signal and gives the $S_{21}$ response. Finally, the high frequency responses of the RF probe and RF amplifier were calibrated to obtain the $S_{21}$ response of the GPD.

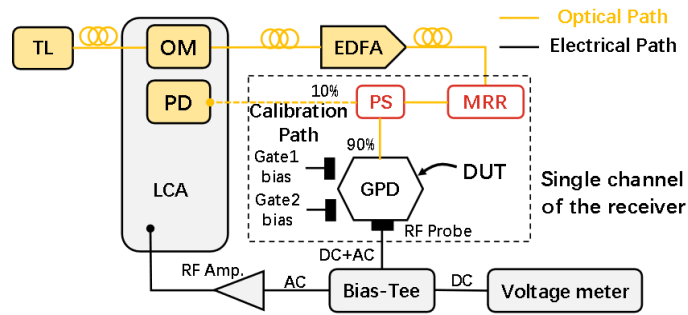

**Figure S6 Detailed experimental set up for the high frequency response measurement.** TL: Tunable laser; EDFA: Erbium-doped fiber amplifier; LCA: Light-wave component analyzer. PS: Power splitter. OM: optical modulator.